\documentclass[a4paper,11pt]{article}

\usepackage{jcappub}
\usepackage{slashed}
\usepackage{bm}
\usepackage{latexsym,amssymb,amsmath,float,url,mathrsfs}
\usepackage{latexsym}
\usepackage{graphicx}
\usepackage{epstopdf}
\usepackage{amsfonts}
\usepackage{amsmath}
\usepackage{amssymb}
\usepackage{comment}
\usepackage{subfigure}
\usepackage{natbib}
\usepackage{hyperref}
\usepackage{pifont}
\usepackage{blindtext}
\usepackage{adjustbox}
\usepackage{multirow}
\usepackage{tabularx}
\usepackage[table]{xcolor}
\usepackage{orcidlink}
\usepackage{tikz}
\usetikzlibrary{tikzmark}
\usepackage{amssymb}
\usepackage{bbold}

\usepackage{blkarray}
\usepackage{graphicx}
\usepackage{amsfonts}
\usepackage{soul}
\usepackage{amssymb}
\usepackage{amsmath}
\usepackage{cancel}
\usepackage{tcolorbox}

\usepackage{graphics,appendix,afterpage,makecell} 

\def\Re{\mbox{Re}\,}

\definecolor{oucrimsonred}{rgb}{0.6, 0.0, 0.0}
\definecolor{persianblue}{rgb}{0.11, 0.22, 0.73}
\definecolor{forestgreen}{rgb}{0.13,0.35,0.13}
\definecolor{lightgray}{rgb}{0.83, 0.83, 0.83}
 \hypersetup{colorlinks, citecolor=oucrimsonred, linkcolor=black, urlcolor=oucrimsonred}
\definecolor{cornellred}{rgb}{0.7, 0.11, 0.11}
\definecolor{navyblue}{rgb}{0.0, 0.0, 0.5}
\definecolor{amethyst}{rgb}{0.6, 0.4, 0.8}
\definecolor{yellow}{rgb}{1.0, 1.0, 0.0}
\definecolor{firebrick}{rgb}{0.7, 0.13, 0.13}
\definecolor{tangerineyellow}{rgb}{1.0, 0.8, 0.0}
\definecolor{deepfuchsia}{rgb}{0.76, 0.33, 0.76}
\definecolor{amber}{rgb}{1.0, 0.75, 0.0}
\definecolor{VioletRed4}{rgb}{0.55, 0.13, .32}
\definecolor{indiagreen}{rgb}{0.07, 0.53, 0.03}
\definecolor{VioletRed4}{rgb}{0.55, 0.13, .32}
\newcommand{\be}{\begin{equation}}
\newcommand{\ee}{\end{equation}}
\newcommand{\bea}{\begin{equation} \begin{aligned}}
\newcommand{\eea}{\end{aligned} \end{equation}}

\definecolor{oucrimsonred}{rgb}{0.6, 0.0, 0.0}
\newcommand\vertarrowbox[3][6ex]{%
  \begin{array}[t]{@{}c@{}} #2 \\
  \left\uparrow\vcenter{\hrule height #1}\right.\kern-\nulldelimiterspace\\
  \makebox[0pt]{\scriptsize#3}
  \end{array}%
}

\definecolor{verdechiaro}{rgb}{0.6,1,0.6}
\definecolor{giallochiaro}{rgb}{1,1,0.6}
\definecolor{bluscuro}{rgb}{0.15, 0.2, 0.9}
\definecolor{verdes}{rgb}{0.1, 0.5, 0.1}%
\definecolor{tangerineyellow}{rgb}{1.0, 0.8, 0.0}

\definecolor{americanrose}{rgb}{1.0, 0.01, 0.24}
\definecolor{cobalt}{rgb}{0.0, 0.28, 0.67}
\definecolor{brandeisblue}{rgb}{0.0, 0.44, 1.0}
\definecolor{mycolor}{rgb}{0.0, 0.0, 0.5}
\definecolor{oxfordblue}{rgb}{0.0, 0.13, 0.28}
\definecolor{azure}{rgb}{0.0, 0.5, 1.0}
\definecolor{turquoiseblue}{rgb}{0.0, 1.0, 0.94}
\newtcolorbox{mynewbox}[1]{colback=white!5!white,colframe=azure!75!black,fonttitle=\bfseries,title=#1}
\newtcolorbox{mybox}{colback=mycolor!5!white,colframe=azure!75!black}
\newtcolorbox{mynamedbox}[1]{colback=mycolor!5!white,colframe=azure!75!black,title=#1}
\definecolor{venetianred}{rgb}{0.78, 0.03, 0.08}
\newtcolorbox{mynamedbox1}[1]{colback=venetianred!5!white,colframe=venetianred!80!black,title=#1}
\newtcolorbox{mynamedbox2}[1]{colback=azure!5!white,colframe=azure!80!black,title=#1}

\newcommand{\td}{{\rm d}}

\definecolor{verdes}{rgb}{0.1, 0.5, 0.1}%
\definecolor{cornellred}{rgb}{0.7, 0.11, 0.11}

\definecolor{VioletRed4}{rgb}{0.55, 0.13, .32}

\hypersetup{
     colorlinks   = true,
     citecolor    = violet,
     urlcolor     = violet,
     linkcolor    = violet}

\definecolor{rossocorsa}{rgb}{0.83, 0.0, 0.0}

\usepackage[normalem]{ulem}

\title{Black Holes in a Gravitational Field:\\
The  Non-linear Static  Love Number  of  Schwarzschild Black Holes Vanishes}

\author[a,b]{A. Kehagias}
\author[b,c]{A. Riotto}
\affiliation[a]{Physics Division, National Technical University of Athens, Athens, 15780, Greece}
\affiliation[b]{Department of Theoretical Physics, 
24 quai E. Ansermet, CH-1211 Geneva 4, Switzerland}

\affiliation[c]{ Gravitational Wave Science Center,  
24 quai E. Ansermet, CH-1211 Geneva 4, Switzerland}

\abstract{
We show that the  static tidal Love number of  Schwarzschild black holes in four dimensions and in the vacuum vanishes at any order in the external tidal force. We  also identify  the  underlying  non-linear symmetry  which  is responsible for this result 
and becomes manifest when the black hole metric is written in axsymmetric static Weyl coordinates.}

\emailAdd{kehagias@central.ntua.gr}
\emailAdd{antonio.riotto@unige.ch}

\begin{document}
\maketitle

\section{Introduction}
\noindent
General Relativity (GR) predicts the existence of Gravitational Waves (GWs)  and Black Holes (BHs) and they are both currently tested through the detection of the GWs generated in BH mergers~\cite{LIGOScientific:2021sio}. If 
the  orbital separation  of the bodies  involved in the early inspiral is sufficiently small,    tidal effects   are relevant and are  described by the   so-called Tidal Love Numbers (TLNs)~\cite{poisson_will_2014}. 

In general, static TLNs  depend on the internal properties and structure of the deformed compact objects and impact the gravitational waveform at the fifth post-Newtonian order~\cite{Flanagan:2007ix}. 
Intriguingly, in four dimensions and in the vacuum, the  static TLNs of non-rotating and spinning BHs are exactly vanishing  at  linear perturbation theory in the external tidal force~\cite{Binnington:2009bb,Damour:2009vw,Damour:2009va,Pani:2015hfa,Pani:2015nua,Porto:2016zng,LeTiec:2020spy, Chia:2020yla,LeTiec:2020bos,Poisson:2021yau,Kehagias:2024yzn}, a result which is possibly related to the appearance   of  underlying hidden symmetries ~\cite{Hui:2020xxx,Charalambous:2021mea,Charalambous:2021kcz,Hui:2021vcv,Hui:2022vbh,Charalambous:2022rre,Ivanov:2022qqt,Katagiri:2022vyz, Bonelli:2021uvf,Kehagias:2022ndy,BenAchour:2022uqo,Berens:2022ebl,DeLuca:2023mio, Rai:2024lho}. 

A simple  way to understand why the static TLN  vanishes at the linear order is to  consider a   massless scalar field $\phi$, as the simplest   proxy for the helicity two tensor degrees of freedom. Expanding the scalar field in spherical harmonics, solving its dynamics in the static case far enough from the BH horizon, the general solution is of the type

\be
\label{a}
\phi_\ell(r\rightarrow \infty)= a_\ell \,r^\ell+\frac{b_\ell}{r^{\ell+1}},
\ee
where $\ell$ is the standard multipole.
The  growing mode $\sim r^\ell$ plays the role of the  the  external  tidal force and the  reaction generated by the body  is tested through  the appearance of  the decaying mode $\sim r^{-\ell-1}$. 
The static TLN  is fixed  by the ratio $b_\ell/a_\ell$ and vanishes if $b_\ell=0$. For non-rotating and spinning BHs this is true  because  the decaying mode  is mapped into the divergent solution at the horizon and therefore must be excluded from the set of physical solutions. 

This simple calculation is extended to the more realistic case in which what is perturbed is the metric.
The linear $\ell=2$ static metric perturbation of a Schwarzschild BH
 in the Regge-Wheeler (RW) gauge and without singularities at the horizon   turns out to be
\begin{align}
    \td s^2=&\Big{(}\overline{g}_{\mu\nu}+\delta g_{\mu\nu}(r,\theta)\Big{)} \td x^\mu\td x^\nu,
\end{align}
where 
\begin{eqnarray}
    \overline{g}_{\mu\nu}={\rm diag}\left[-\left(1-\frac{r_s}{r}\right),\dfrac{1}{1-\frac{r_s}{r}}, r^2,r^2 \sin^2\theta\right],
\end{eqnarray}
$r_s=2G_N m$ is the Schwarzschild in terms of the mass $m$ of the BH, and (for the even-parity modes)
   \begin{align}
    \delta g_{\mu\nu}=\mathcal{E}_m{\rm diag}&\left[
    \frac{r^2}{r_s^2}\left(1-\frac{r_s}{r}\right)^2,\, 
    \frac{r^2}{r_s^2}, \, \frac{r^4}{r_s^2}\left(1-\frac{r^2_s}{r^2}\right),\nonumber\right.\\
    &\left.\frac{r^4}{r_s^2}\left(1-\frac{r^2_s}{r^2}\right)\sin^2\theta\right]Y_{2 m}(\theta,\phi),
    \label{dg}
\end{align}
where $\mathcal{E}_m$ is the amplitude of the tidal force.
The absence of a decaying $\sim r^{-3}$ mode, due the fact that  it is mapped into a divergence at the horizon, shows that the corresponding TLN is zero at linear order.
The perturbed, second-order in $\mathcal{E}_m$, Einstein equations have been recently solved in Ref. \cite{Riva:2023rcm,Riva:2024}, where the explicit solution is given in the RW gauge. There is no decaying mode and again the static TLN vanishes at this order.

These results raise a simple, but fundamental question: is the static TLN of Schwarzschild BHs vanishing at any order in perturbation theory? Answering this question goes beyond a purely academic interest. For instance, 
a zero static TLN for Schwarzschild and Kerr BHs would be  crucial to    distinguish BHs from neutron stars in sub-solar mergers, thus eventually confirming  the primordial nature of the BHs \cite{Crescimbeni:2024cwh,Riotto:2024ayo}. 

In this paper we  will show that the answer is yes, the static TLN for Schwarzschild BHs vanishes at any order in perturbation theory in the  external tidal force. We will also show that the underlying reason is  again a ladder   symmetry manifestly popping out when choosing a suitable set of coordinates.

The paper is organized as follows. In section 2 we offer the proof, reproducing as well the standard linear result in the RW gauge;  in section 3 we identify the underlying non-linear symmetries and we conclude in section 4.

\vskip 0.3cm
\noindent
{\bf Note Added} After the completion of this work,  we have become aware of a similar work by O. Combaluzier-Szteinsznaider, L. Hui, L. Santoni, A. R. Solomon, and S. Wong entitled ``Symmetries of Vanishing Nonlinear Love Numbers of Schwarzschild Black Holes'' and have coordinated the submission to the archive.  Their results, when overlap is possible, agree with ours.

\section{The proof for the Schwarzschild BH}
\noindent
The underlying idea leading to a non-perturbative statement is to 
rephrase the problem in a suitable way such that  perturbation theory is considerably simplified and the  symmetries of the system more manifest. 

 Let us consider therefore a static axisymmetric vacuum spacetime in Weyl coordinates $(t,\rho,z,\phi)$ (the preferred direction being the axis connecting the two orbiting BHs)  with metric~\cite{Papapetrou:1953zz}
	\begin{eqnarray}
	\td s^2&=&-e^{2 U}(\td t+\omega \td \phi)^2
	+e^{-2U}e^{2k}\left(\td \rho^2+\td z^2\right)
 +\rho^2 e^{-2U}\td \phi^2, \label{ds}
	\end{eqnarray}
	where $U=U(\rho,z)$ and $k=k(\rho,z)$. We will not consider rotation   from now on, setting $\omega=0$.  The vacuum  Einstein equations  $R_{\mu\nu}=0$ are 
	\begin{eqnarray}
	&&\nabla^2U=
 \partial_{\rho\rho}U
	+\frac{1}{\rho}\partial_\rho U+\partial_{zz}U=0, 
 \label{dU}\\
	&&\partial_\rho k=\rho\left[(\partial_\rho U)^2+(\partial_z U)^2\right],\,\, \partial_zk=2\rho\,  \partial_\rho U \partial_z U. \label{dk}\nonumber\\
 &&
	\end{eqnarray}
 Interestingly enough, the function $U$ satisfies a linear Laplace equation whereas the non-linearities of the Einstein equations are all encoded in the first-order  partial differential equation obeyed by $k$.  This will play a crucial role when explaining the vanishing of the TLN through symmetries. In addition, it can easily be checked that there is always a solution for $k$ once a solution of the Laplace equation Eq. (\ref{dU}) for $U$ on $\mathbb{R}^3$ is given as the integrability condition ${\rm d}k=0$ for Eq. (\ref{dk}) is satisfied.
 Note also that regularity at the axis requires $k(\rho\to 0)\to 0$. 
 
There is some merit in using prolate spheroidal coordinates $(t,x,y,\phi)$ instead of Weyl  coordinates \cite{Zipoy:1966btu,Ernst:1967wx}
 by writing 
 \begin{eqnarray}
     \rho&=&\rho_0 (x^2-1)^{1/2}(1-y^2)^{1/2}, \qquad
     x\geq 1, \qquad |y|\leq 1, \nonumber \\
     z&=&\rho_0 x y, \qquad \rho_0=\mbox{constant.}
 \end{eqnarray}
 Regularity on the axis now means that 
 $k(y^2\to 1)\to 0$.
 In such coordinates, the metric in Eq. (\ref{ds}) is written as 
 \begin{eqnarray}
     \td s^2&=&-e^{2U(x,y)}\td t^2+
    e^{-2U(x,y)}\left[e^{2k(x,y)}\rho_0^2 (x^2-y^2)\left(\frac{\td x^2}{x^2-1}+\frac{\td y^2}{1-y^2}\right)
    +\rho_0^2 (x^2\!-\!1) (1\!-\!y^2)\td \phi^2\right]\nonumber\\
    &&
     \label{dsp}
 \end{eqnarray}
and  Einstein equations become

\begin{eqnarray}
    &&\bigg[  \partial_x(x^2-1)\partial_x+\partial_y(1-y^2)\partial_y \bigg]U(x,y)=0, \label{dUx}\\
    &&\partial_x k=\frac{1-y^2}{x^2-y^2} \bigg[
    x(x^2-1)(\partial_x U)^2-2 y(x^2-1) \partial_x U\partial_y U-x(1-y^2) (\partial_y U)^2\bigg], \label{dkx}\\
    &&\partial_y k=\frac{x^2-1}{x^2-y^2} \bigg[
    y(x^2-1)(\partial_x U)^2+2 x(1-y^2) \partial_x U\partial_y U-y(1-y^2) (\partial_y U)^2\bigg]. \label{dky}
\end{eqnarray}
By expressing the solution of Eq. (\ref{dUx}) as \cite{Ernst:1967wx} 
\begin{eqnarray}
    U(x,y)=\sum_{\ell} U_\ell(x) Y_\ell(y),
\end{eqnarray}
we find that $U_\ell$ and $Y_\ell$
satisfy 
\begin{eqnarray}
    &&\frac{\td}{\td x}\left((x^2-1)\frac{\td}{\td x}U_\ell\right)-\ell(\ell+1)U_\ell=0, \label{Xl}\\
    &&\frac{\td}{\td y}\left((1-y^2)\frac{\td}{\td y}Y_\ell\right)+\ell(\ell+1)Y_\ell=0. \label{Yl}
\end{eqnarray}
The solution of Eq. (\ref{Yl}) which is regular at $y=\pm 1$ is given by the Legendre polynomials 
\begin{eqnarray}
    Y_\ell(y)=P_\ell(y), \qquad \ell=0,1,\cdots. 
\end{eqnarray}
Similarly, the solution to Eq. (\ref{Xl}) is 
\begin{eqnarray}
    U_\ell&=&\alpha_\ell \, x^\ell \, {}_2F_1\left(\frac{1-\ell}{2},-\frac{\ell}{2},\frac{1-2\ell}{2},\frac{1}{x^2}\right)+\beta_\ell \frac{1}{x^{\ell+1}}\, {}_2F_1\left(\frac{1+\ell}{2},\frac{2+\ell}{2},\frac{3+2\ell}{2},\frac{1}{x^2}\right), \nonumber\\
    &&
    \label{fund}
\end{eqnarray}
so that the function $U(x,y)$ turns out to be
\begin{eqnarray}
    U(x,y)&=&\sum_{\ell=0}^\infty \bigg[\alpha_\ell \, x^\ell \, {}_2F_1\left(\frac{1\!-\!\ell}{2},-\frac{\ell}{2},\frac{1\!-\!2\ell}{2},\frac{1}{x^2}\right)+\frac{\beta_\ell}{x^{\ell+1}}\, {}_2F_1\left(\frac{1\!+\!\ell}{2},\frac{2\!+\!\ell}{2},\frac{3\!+\!2\ell}{2},\frac{1}{x^2}\right)\bigg] P_\ell(y). \nonumber\\
    &&
    \label{Uf}
\end{eqnarray}
By using the properties of the hypergeometric functions, it is easy to see  that the growing mode proportional to $\alpha_\ell$ is a polynomial of $x$ of degree $\ell$. On the other hand, the decaying mode proportional to $\beta_\ell$ is singular at $x=1$. This point  should be excluded from this mode since the associated hypergeometric function does not converge there.

As we have mentioned above, once the function $U(x,y)$ is specified, then $k(x,y)$ can be found by solving Eqs. (\ref{dkx}) and (\ref{dky}). However, although $k(x,y)$ could explicitly be specified in the general case,   we will restrict ourselves in some special cases to get some more intuitive insights into our findings.

\subsection{The monopole: recovering Schwarzschild solution}
\noindent
Let us consider first the monopole $\ell=0$ mode \cite{Ernst:1967wx}.  In this case Eq. (\ref{Uf}) gives  
\begin{eqnarray}
    U=\alpha_0+
    \frac{\beta_0}{2}\ln\left(\frac{x-1}{x+1}\right),
\end{eqnarray}
and using  the above expression in  Eqs. (\ref{dkx}) and (\ref{dky}), we find that 
\begin{eqnarray}
    k=\frac{\beta_0^2}{2}\ln\left(\frac{x^2-1}{x^2-y^2}\right) 
\end{eqnarray}
by imposing  regularity on the axis of symmetry, that is $k(y^2\to 1)\to 0$. In addition, there are potential singularities, which can be 
 revealed by examining  geometric scalars like  the Kretschmann scalar invariant
\begin{eqnarray}
    K=R_{\mu\nu\rho\sigma}\,
    R^{\mu\nu\rho\sigma}. \label{K}
\end{eqnarray}
The latter becomes
\begin{align}
    K= &\frac{f(x,y)}{m^4} \frac{\beta_0 (1-\beta_0)^2(1-\beta_0+\beta_0^2)}{(1-x)^{2(1+\beta_0+\beta_0^2)}(1+x)^{2(1-\beta+\beta^2)}},
\end{align}
where $f(x,y)$ is a polynomial in $x$ and $y$.  
It can easily be verified that there is always a singularity at $x=1$. The only possibility is either $\beta_0=0$, or $\beta_0=1$. The former case corresponds to flat Minkowski spacetime and the
latter to Schwarzschild. Therefore, we will assume that $\beta_0=1$ so that the metric (\ref{dsp}) is 
\begin{eqnarray}
    \td s^2&=&-\frac{x-1}{x+1}\td t^2+\frac{x+1}{x-1}
    \rho_0^2 \td x^2+(x+1)^2 \rho_0^2 \frac{\td y^2}{1-y^2}+\rho_0^2 (x+1)^2(1-y^2)\td \phi^2. 
    \label{ds11}
\end{eqnarray}
We have also put $\alpha_0=0$ as a constant shift in the function $U$ can be absorbed in a constant rescaling of the coordinates. Note that there is a horizon for the metric 
in Eq. (\ref{ds11}) at $x=1.$
By changing now coordinates and write 
\begin{eqnarray}
    x&=&2r/r_s-1, \qquad \rho_0=r_s/2, \nonumber \\
    y&=&\cos\theta, \,\,\,\,\,\,\,\,\,\,\,\,\,\,\,\,\,\,\, \,\,\,\,\,0\leq \theta\leq \pi,
    \label{coord}
\end{eqnarray}
so that the horizon is at $r=r_s$ as usual, 
the metric Eq. (\ref{ds11}) is written as
\begin{align}
    \td s^2=&-\left(1-\frac{r_s}{r}\right)\td t^2
    +\dfrac{\td r^2}{1-\frac{r_s}{r}}+r^2 \td \theta^2
    +r^2 \sin^2\theta\td \phi^2,
\end{align}
which is,  the standard form of the Schwarzschild metric for a BH and horizon at $r_s$.

\subsection{The monopole and quadrupole: tidal forces}
\noindent
This case describes a BH immersed in a quadruple gravitational field. The function $U$ is now
\begin{eqnarray}
    U(x,y)&=&\frac{1}{2}\ln\left(\frac{x-1}{x+1}\right)+\left[ A \left(x^2-\frac{1}{3}\right)+\frac{B}{x^3}
    {}_2F_1\left(\frac{3}{2},2,\frac{7}{2},\frac{1}{x^2}\right)\right]\left(\frac{1}{2}-\frac{3}{2}y^2\right), 
    \label{U2}
\end{eqnarray}
where we have redefined $\alpha_2=A$ and $\beta_2=B$.
Using the expression (\ref{U2}) above in Eqs. (\ref{dkx}) and (\ref{dky}) and integrating, we end up in a long expression for $k$, which nevertheless is not particularly illuminating. 

What we wish  to do is instead  to examine if the two modes of $\ell=2$, the growing and the decaying ones proportional to $A$ and $B$ respectively, are physically acceptable. This can be answered by examining possible singularities of the underlying spacetime. 
We can again consider the 
the Kretschmann scalar in Eq. (\ref{K})
 close to the potential problematic point $x=1$
 Expanding Eq. (\ref{U2}), integrating Eqs. (\ref{dkx}) and (\ref{dky}) and then calculating (\ref{K}), we find  
\begin{align}
    K=&\frac{2304}{r_s^4(1+x)^6}-4\frac{A}{r_s^4}(29+9\cos2\theta)+\cdots-B\bigg[
    \frac{1}{ r_s^4 (1-y^2)}+\ln(x-1)+\cdots\bigg].
\end{align}
Therefore, for any non-vanishing $B$ the spacetime is singular at $x=1$ and $y=\pm 1$. Thus, we must choose $B=0$ to avoid a singular  spacetime at the horizon from the decaying mode.  In this case, the function $U$ becomes 
\begin{eqnarray}
    U(x,y)=\frac{1}{2}\ln\left(\frac{x-1}{x+1}\right)+ A \left(x^2-\frac{1}{3}\right)\left(\frac{3y^2}{2}-\frac{1}{2}\right). 
    \nonumber\\
    &&\label{U22}
\end{eqnarray}
 Integrating  Eqs. (\ref{dkx}) and (\ref{dky}) we find that 
 \begin{eqnarray}
     k(x,y)&=&\frac{1}{2}\ln\left(\frac{x^2-1}{x^2-y^2}\right)-2 A x (1-y^2)+
    \frac{A^2}{4} \bigg[y^4-2 y^2-2 x^2 (1-6 y^2 +5 y^4)+x^4 (1-10 y^2 +9 y^4)\bigg]\nonumber\\ 
    &=&-\frac{4 x}{3}  A-\frac{A^2}{60}(7+8 x^4)+\left[\frac{4 x}{3}  A-\frac{4 A^2}{21}(1-3 x^2+2 x^4)\right] P_2(y)\nonumber\\
    &+&\frac{2A^2}{35}(1-10x^2+9 x^4) P_4(y)+\frac{1}{2}\ln\left(\frac{x^2-1}{x^2-y^2}\right)
 \end{eqnarray}
and the metric (\ref{dsp}) turns out to be in the coordinates $(r,\theta)$ of Eq. (\ref{coord})
\begin{eqnarray}
    \td s^2&=&
    e^{-A(2/3 -2\frac{r}{2 r_s}+r^2/m^2)(3 \cos\theta^2-1)}\left[
    e^{2 k_0(r,\theta)}
      \left(\dfrac{\td r^2}{1-\frac{r_s}{r}}+r^2 \td \theta^2\right)+r^2 \sin^2\theta \td \phi^2\right]\nonumber\\
    &-&\left(1-\frac{r_s}{r}\right)
    e^{A(2/3 -2\frac{r}{2 r_s}+r^2/m^2)(3 \cos\theta^2-1)}\td t^2,
    \label{ds22}
\end{eqnarray}
where $k_0(r,\theta)$ is given by 
\begin{eqnarray}
   k_0(r,\theta)&=& -2 A \left(\frac{r}{2 r_s}-1\right) \sin^2\theta+
    \frac{A^2}{4} \bigg[\cos^4\theta-2 \cos^2\theta
    -2 \left(\frac{r}{2 r_s}-1\right)^2 (1-6 \cos^2\theta +5 \cos^4\theta)\nonumber \\
&+&\left(\frac{r}{2 r_s}-1\right)^4 (1-10 \cos^2\theta +9 \cos^4\theta)\bigg]. 
\end{eqnarray}
Again, there is a horizon at $x=1$ ($r=r_s)$. 
Close to the horizon, the Kretschmann scalar $K$ defined in Eq. (\ref{K}) for the metric (\ref{ds22})  is finite at $r=r_s$ as expected
\begin{align}
K=&\frac{12}{r_s^4} e^{A(20+3A-6\cos2\theta)/3} \bigg[ 1-A(3+A)-
5 A \cos2\theta+A^2 \cos4\theta
\bigg]^2 +\mathcal{O}(r-r_s).    
\end{align}
Eq. (\ref{ds22}) is our main result. It describes a BH embedded in a quadruple gravitational environment. There is no decaying $\ell=2$ mode in  the full exact solution (\ref{ds22}). The TLN is vanishing at any order in the external tidal force\footnote{We remark that in Ref. \cite{Gurlebeck:2015xpa}  only Geroch multipole moments \cite{Geroch:1970cd} have been considered. The latter are defined for  asymptotically flat  spacertimes and therefore are relevant for no-hair theorems and not for the problem at hand where the BH is immersed in a gravitational field scaling like $\sim r^\ell$.}. 

\subsection{Going back to the linear level and comparison with the result in the RW gauge} 
\noindent
Let us expand the solution (\ref{ds22}) in powers of $A$ for $r_s\ll r\ll r_s/\sqrt{A}$, to
 compare our results with the existing literature at the linear level.  To first-order the metric reads    
\begin{align}
    \td s^2=&\Big{(}\overline{g}_{\mu\nu}+h_{\mu\nu}(r,\theta)\Big{)} \td x^\mu\td x^\nu\nonumber,
\end{align}
where again $\overline{g}_{\mu\nu}$ is the background Schwarzschild metric and  $h_{\mu\nu}$ is the first-order perturbation. We take  first-order perturbation of the form 
\begin{eqnarray}
    h_{\mu\nu}=h_{0\mu\nu}(r,\theta)
    +h_{2\mu\nu}(r,\theta)P_2(\cos\theta). 
\end{eqnarray}
where $h_{0\mu\nu}$ and $h_{2\mu\nu}$ are the $\ell=0$ and $\ell=2$ moments, respectively. Expanding in first-order in $A$, they are
 explicitly given by
\begin{align}
    h_{a\mu\nu}={\rm{diag}}\bigg(h_{a0},h_{a1},h_{a2},h_{a3}\bigg),\,\,  a=0,2,
\end{align}
where 
\begin{eqnarray}
    h_{00}&=&0,\\
    h_{01}&=&-\frac{32 A r}{3r_s}\frac{r-r_s/2}{r-r_s},\\
    h_{02}&=&-\frac{32 A r^3}{3r_s}\left(1-\frac{r_s}{2r}\right), \\
    h_{03}&=&-\frac{8}{45}A r^2 \left[1-3\left(1-\frac{2r}{r_s}\right)^2\right]\\
    h_{20}&=&\frac{2A}{3}\left(1-\frac{r_s}{r}\right)\left[1-3 \left(1-\frac{2r}{r_s}\right)^2\right],\\
    h_{21}&=&-\frac{8 A }{3 r_s^2}\dfrac{1}{1-\frac{r_s}{r}}\bigg(3 r_s^2/2-5 r_s r+3 r^2\bigg), \\
    h_{22}&=&-\frac{8 A r^2 }{3r_s^2} \bigg(3r_s^2/2-5 r_s r+3 r^2\bigg),\\
    h_{23}
    &=&-\frac{20}{63}A r^2 \left[1-3\left(1-\frac{2r}{r_s}\right)^2\right].
\end{eqnarray}
The  gravitational perturbation $h_{\mu\nu}$  can be brought to
RW gauge by a gauge transformation generated by the vector $\xi^\mu$
\begin{eqnarray}
    h_{\mu\nu}^{\rm RW}=h_{\mu\nu}-\overline{\nabla}_\mu\xi_\nu-
    \overline{\nabla}_\nu\xi_\nu,
\end{eqnarray}
where $\overline{\nabla}_{\mu}$ is the covariant derivative with respect to the background metric $\overline{g}_{\mu\nu}$. 
The explicit form of $\xi^\mu$ is 
\begin{eqnarray}
    \xi^\mu&=&\left[-\xi_0(t,r),\, \frac{2A r}{3r_s}\Big{(}
    2 r_s -3r_s \cos^2\theta-3r \sin^2\theta\Big{)},
    -A\left(1-\frac{2r}{r_s}\right)\sin2\theta,\, \xi_3\right],
\end{eqnarray}
where $\xi_3=\mbox{const.}$ Then, the metric perturbation in the RW gauge turns out to be
\begin{eqnarray}
    h_{\mu\nu}^{\rm RW}=h_{0\mu\nu}^{\rm RW}+h_{2\mu\nu}^{\rm RW}P_2(\cos\theta),
\end{eqnarray}
where 
\renewcommand{\arraystretch}{1.4}
\begin{eqnarray}
    h_{000}^{\rm RW}&=&
-\dfrac{8A}{3}\left(1\!-\!\dfrac{r_s}{2r}\right)-4\left(1\!-\!\dfrac{r_s}{r}\right) \dot{\xi_0}\\
 h_{001}^{\rm RW}&=&-2\left(1-\dfrac{r_s}{r}\right)\xi_0'\\
 h_{011}^{\rm RW}&=&-\dfrac{4Ar_s}{3r}\dfrac{1}{\left(1-\dfrac{r_s}{r}\right)^2},
\end{eqnarray}
and all the other components vanishing, and 
\begin{eqnarray}
    h_{2\mu\nu}^{\rm RW}&=&-8A\,{\rm diag}\left[
     \frac{r^2}{r^2_s}\left(1-\frac{r_s}{r}\right)^2,
     \frac{r^2}{r^2_s},\frac{r^4}{r^2_s}\left(1-\frac{r^2_s}{r^2}\right),\frac{r^4}{r^2_s}\left(1-\frac{r^2_s}{r^2}\right)\sin^2\theta\right]. \nonumber\\
    &&
    \label{h2RW}
\end{eqnarray}
With the parametrization 
\begin{eqnarray}
    A=-\frac{1}{16}\sqrt{\frac{5}{\pi}}\mathcal{E}_m. 
\end{eqnarray}
  Eq. (\ref{h2RW}) coincides with Eq. (\ref{dg}). As an aside technical point, we stress that although one can solve the perturbative equations for $\ell=2$ only, we see  that  turning on the $\ell=2$ mode in the exact solution (\ref{ds22}) generates also an $\ell=0$ mode in the perturbative expansion. This is expected as  in the full non-linear solution the various  multipoles are interacting among themselves.

\section{Non-linear symmetries}
\noindent
As mentioned in the introduction,  in four dimensions and in the vacuum, the  static TLNs of non-rotating and spinning BHs are exactly vanishing  at  linear perturbation theory because of the  appearance   of  underlying hidden symmetries in the linear equations of motion~\cite{Hui:2020xxx,Charalambous:2021mea,Charalambous:2021kcz,Hui:2021vcv,Hui:2022vbh,Charalambous:2022rre,Ivanov:2022qqt,Katagiri:2022vyz, Bonelli:2021uvf,Kehagias:2022ndy,BenAchour:2022uqo,Berens:2022ebl,DeLuca:2023mio, Rai:2024lho}. The underlying idea is that, for each mode $\ell$ solving the linear equation of motion, there is a symmetry and conserved charge $P_\ell$.  One can  descend to the mode $\ell = 0$ by a series of lowering operations, and then invoke the conservation of $P_0$ to see that $P_\ell$ is conserved. Its utility resides on the fact that  from it one can   understand why the decaying solution $\sim 1/r^{\ell+1}$ is directly connected to the divergent solution at the horizon and therefore has to be disregarded, leading to an argument for the nonvanishing TLN.

This raises the question if there is a non-linear underlying symmetry explaining the vanishing of the static TLN, extending the linear argument. We now show  that it is the case.

 The crucial  remark is that   Eq. (\ref{Xl}), despite the full non-linear problem, is a linear equation and is exactly the same equation considered in the literature addressing the linear case  for a static, massless scalar on a Schwarzschild background.  
 
 On the contrary, the full non-linearities are encoded in the function $k(x,y)$ which however is fully determined once the function $U(x,y)$ is known. To proceed, 
we  define ladder operators 
\begin{align}
    L^+_\ell=&-(x^2-1)\frac{\td}{\td x}-(\ell+1) x,\nonumber \\
    L^-_\ell=&(x^2-1)\frac{\td}{\td x}-\ell x,
    \label{lad}
\end{align}
such that $L^\pm_\ell$ are multipole raising/lowering operators

\begin{eqnarray}
  L^+_\ell U_\ell\sim U_{\ell+1},\,\,\,\,\,\,   L^-_\ell U_\ell\sim U_{\ell-1}.
\end{eqnarray}
Following the standard construction done at the linear level \cite{Hui:2021vcv}, one can show that there are  conserved quantities

\begin{equation}
    P_\ell=(x^2-1)\frac{\td}{\td x}\left(L_1^- L_2^-\cdots L_\ell^-\right)U_\ell,
\end{equation}
for which

\begin{equation}
    \frac{\td P_\ell}{\td x}=0.
\end{equation}
To show that the decaying in Eq. (\ref{fund}) has to be disregarded, we consider the large $x$ limit for which the decaying mode goes like 

\begin{equation}
U_\ell \sim\frac{\beta_\ell}{x^{\ell+1}}.
\end{equation}
The corresponding conserved $P_\ell$ is finite and not zero at large $x$. Close to the horizon such a decaying mode  diverges as $\ln(x-1)$. 
Since near the horizon the growing and the decaying modes  must have the same corresponding $P_\ell$ they have at infinity and since the growing mode close the horizon is   constant and therefore has $P_\ell = 0$,  the conservation of $P_\ell$ tells us the decaying solution is the one   diverging logarithmically at the horizon.
However, this is not enough yet to conclude that the decaying mode has to be disregarded since the logarithmic piece is there to reproduce exactly the monopole Schwarzschild background in Schwarzschild coordinates. On the contrary, the decaying mode in $U(x,y)$ has to be thrown away because it gives an extra logarithmic divergent contribution to the  Kretschmann scalar. No extra decaying modes are produced in the non-linear function $k(x,y)$ once the decaying modes are excluded in the function $U(x,y)$.

As a final remark, we notice that the Laplacian equation (\ref{dU}) is the same equation in a two-dimensional flat spacetime in cylindric coordinates. 
As such, its solutions can always be expressed on complex coordinates $(\zeta,\bar\zeta)$ as the sum a holomorphic 
function $f(\zeta)$ plus its complex conjugate. Therefore any analytic holomorphic transformation of $\zeta$ will produce a new solution. 
The  ladder operators are expected to part of the generators of such  conformal (holomorphic) infinite  group of transformations. In fact, 
stationary axisymmetric spacetimes have a rich symmetry structure originating from the two-dimensional aspect of the theory in this case.  The metric 
(\ref{ds}) (with $\omega=0$) has two abelian Killing vectors $\partial_t$ and $\partial_\phi$, and can be written as
\begin{eqnarray}
    \td s^2=f(\zeta,\bar{\zeta})\td \zeta \td \bar{\zeta}+g_{ab}(\zeta,\bar{\zeta})\td x^a\td x^b,\qquad a,b=1,2,
\end{eqnarray}
where $\zeta=\rho+i z$ and $x^1=t$ and $x^2=\phi.$ Then,  Einstein equations for the $2\times 2$ matrix  $g=(g_{ab})$
are 
\begin{eqnarray}
    \partial\left(\sqrt{\det{g}} g^{-1}\bar{\partial}g\right)+
    \bar{\partial}\left(\sqrt{-\det{g}} g^{-1}\partial g\right)=0.  \label{g-1}. \label{s-g}
\end{eqnarray}
 It can be easily verified that this is the   equations of
motion an SL$(2,\mathbb{R})$/U(1)  non–linear $\sigma$–models in two dimensions, with an extra $\sqrt{-{\rm det}\,{g}}$ factor. Non-linear $\sigma$–models of this type are encountered often in general relativity and  
called, in general, Ernst models. They appear  in  Geroch’s  reduced gravity approach \cite{Geroch:1972yt} and have been scrutinized 
 by Ernst \cite{Ernst:1967wx}. 
Eq. (\ref{s-g}) exhibits the symmetry
\begin{eqnarray}
    \delta_T g=-\frac{1}{s}\Re\bigg(F(s) TF(s)^{-1}i \epsilon\bigg), 
\end{eqnarray}
where $s$ is a spectral parameter, 
$T=\epsilon_i T_i$ is a generic element  of SL$(2,\mathbb{R})$ with parameters $\epsilon_i$  ($i=1,2,3$), generated by $T_i$ and $F(s)$ is a $2\times 2$ matrix such that $F(0)=1$. Expanding the operator $\delta_T $ as 
\begin{eqnarray}
    \delta_T=\sum_{n=0}^\infty s^n \delta_T^{(n)},
\end{eqnarray}
one finds that $\delta_T^{(n)}$ satisfy 
\begin{eqnarray}[\delta_{T_1}^{(n)},\delta_{T_2}^{(m)}]=\delta_{T_1+T_2}^{(m+n)}, 
\end{eqnarray}
which is an {SL}(2,$\mathbb{R})$ infinite dimensional current algebra. All symmetry operators, including the ladder operators, are contained in this algebra, which is originating from the two-dimensional nature of the theory. The group actins on the   space of stationary, axially symmetric solutions of
Einstein’s vacuum field equations, producing other solutions \cite{Kerns,Ernst}.

\section{Considerations on the gauge invariance and conclusions}
\noindent
We have offered a proof  of why the static TLN of spinless BHs is vanishing in the vacuum and in four dimensionial general relativity  at any order in the external tidal force. We have shown it by constructing the fully 
non-perturbative solution of Einstein equations showing that no decaying mode is present and we have also identified the ladder symmetry preventing the static TLN from being nonvanishing.

We notice that, differently from the mass and the spin of BHs, which are conserved charges and gauge invariant objects, even though the definition of the TLNs is unambiguous at the Newtonian level \cite{Hinderer:2007mb},
the  static TLN is not a conserved charge and is coordinate dependent in general relativity (see, for instance, Ref. \cite{Gralla:2017djj,Binnington:2009bb,Katagiri:2023umb,Katagiri:2024wbg}).

This has motivated the definition of the linear static TLN   as the Wilson coefficient matching the appropriate operator in the worldline effective action. However, even this operation requires fixing a gauge (the de Donder gauge to simplify expression of the graviton propagator)  and therefore going from the de Donder gauge to whatever gauge is used (the RW gauge at the linear level) to perform the matching. This operation, at the non-linear level might be a difficult task.

Of course, it would be convenient  to define a gauge invariant expression of the TLN, but even this procedure  is not uniquely defined as one can define an infinite number of gauge invariant quantities starting from the expression in a given gauge.  Another option is to fix a suitable gauge. 
Which one  is a matter that can only be decided once the specifics of the measurement are understood. For instance, in the   cosmological setting the halo bias parameter is better defined in the synchronous coordinates  used in the   spherical collapse model \cite{Wands:2009ex}. The problem is that the static TLN  is not directly measured, but extracted from the data through  Bayesian analysis starting from some fitting model-dependent  waveform.

A third option is to start from a scalar quantity, like the Weyl scalar $\Psi_4$ whose perturbation is gauge invariant at the linear level (since its background value vanishes), but this property is already lost at second-order of perturbation theory.

Admittedly, our demonstration is  coordinate dependent, but 
it follows what is routinely done in the literature, that is to  identify the  value of the static TLN through the    analytic properties of hypergeometric functions and the coefficient of the  decaying mode. Our procedure has also the virtue of being fully non-perturbative 
and the   vanishing of the Love number  is  based on 
inspecting the behaviour of the Kretschmann scalar next to the horizon. This statement is therefore  fully coordinate independent.

On a more ahead note,  it would be also interesting to consider the case of spinning BHs, generalizing our results to the Kerr metric, maybe using the zero angular momentum observer frame which  co-rotates with the same angular speed around the BH.  Furthermore, it would be intriguing to  confirm our findings by a simpler formulation  of Einstein’s general relativity with only cubic interactions and performing the matching with the  worldline effective action \cite{mougiakakos:2024nku}. We leave these issues for a forthcoming publication \cite{inprep}.

\begin{acknowledgments}
\noindent
 We thank V. Cardoso, V. De Luca,  P. Pani, and P. Serpico,  for interesting discussions and feedbacks on the draft.  A.R.  acknowledges support from the  Swiss National Science Foundation (project number CRSII5\_213497)
and by  the Boninchi Foundation for the project ``PBHs in the Era of GW Astronomy''. A.K. acknowledges support from the  Swiss National Science Foundation  (project 
number  IZSEZ0\_229414).
\end{acknowledgments}

\bibliographystyle{JHEP}
\bibliography{draftJCAP}
\end{document}